\documentclass[letter]{aa} 
\usepackage{graphicx}
\usepackage{txfonts}
\usepackage{natbib}
\bibpunct{(}{)}{;}{a}{}{,}
\begin{document}
   \title{Indications for 3\,Mpc-scale large-scale structure associated with \\ 
   an X-ray luminous cluster of galaxies at z=0.95\thanks{Based on observations obtained with ESO Telescopes at the Paranal Observatory under programme ID 72.A-0706 and 275.A-5059, and observations collected at the Centro Astron\'omico Hispano Alem\'an (CAHA) at Calar Alto, operated jointly by the Max-Planck Institut f\"ur Astronomie and the Instituto de Astrof\'isica de Andaluc\'ia (CSIC).}}



   \author{R. Fassbender
          \inst{1}
          \and
          H. B\"ohringer
          \inst{1}
          \and
          G. Lamer
          \inst{2}
          \and
          C.R. Mullis
          \inst{3}
          \and
          P. Rosati
          \inst{4}
          \and
          A. Schwope
          \inst{2}
          \and
          J. Kohnert
          \inst{2}
          \and
          J.S. Santos
          \inst{1}
          }


   \institute{Max-Planck-Institut f\"ur extraterrestrische Physik (MPE),
              Giessenbachstrasse~1, D-85748 Garching, Germany \\
              \email{rfassben@mpe.mpg.de}
         \and
            Astrophysikalisches Institut Potsdam (AIP),
            An der Sternwarte~16, D-14482 Potsdam, Germany
         \and       
         University of Michigan, Department of Astronomy, 918 Dennison Building, Ann Arbor, MI 48109-1090, USA      
         \and
         European Southern Observatory (ESO), Karl-Scharzschild-Str.~2, D-85748 Garching, Germany   
             }


   \date{Received 6 November 2007; accepted 24 January 2008}

 
  \abstract
   {X-ray luminous clusters of galaxies at $z\!\sim\!1$ are emerging as major cosmological probes and are fundamental tools to study the cosmic large-scale structure and environmental effects of galaxy evolution at large look-back times.}
   {We present details of the newly-discovered galaxy cluster XMMU\,J0104.4-0630 at $z\!=\!0.947$ and a probable associated system in the LSS environment.}
   {The clusters were found in a systematic study for high-redshift systems using  deep archival XMM-Newton data for the serendipitous  detection and the X-ray analysis, complemented by optical/near-infrared (NIR) imaging observations and spectroscopy of the main cluster.}
   {We find a well-evolved, intermediate luminosity cluster with $L^{\mathrm{0.5-2.0\,keV}}_{\mathrm{X}}\!=\!(6.4\pm 1.3) \times 10^{43}\,h^{-2}_{70}$\,erg\,s$^{-1}$ 
    and strong central 1.4\,GHz radio emission.
   The cluster galaxy population exhibits a pronounced transition toward bluer colors at cluster-centric distances of 1--2 core radii, consistent with an age difference of 1--2\,Gyr for a single burst solar metallicity model.
   The second, less evolved X-ray cluster at a projected distance of 6.4\arcmin \, ($\sim$3\,Mpc) and a concordant red-sequence color likely forms 
    a cluster-cluster bridge with the main target as part of its surrounding large-scale structure at $z\!\simeq\!0.95$.
     }
  {}


   \titlerunning{3\,Mpc-scale LSS associated with an X-ray luminous cluster of galaxies at z=0.95}
   \authorrunning{R. Fassbender et al.}

   \maketitle
%

\section{Introduction}
Distant, X-ray luminous clusters of galaxies have recently gained significant interest for their potential as sensitive cosmological probes, in particular, for future Dark Energy studies. This interest is reflected in several upcoming major surveys in X-rays \citep[eROSITA, ][]{Predehl2006} and the SZ-effect 
\citep[e.g., the South Pole Telescope, ][]{Carlstrom2006}.




XMM-Newton with its high sensitivity and 30\arcmin \, 
field-of-view (FoV) is currently the most efficient X-ray observatory to detect rare, distant X-ray luminous clusters. $z\!\ga\!0.9$ systems are now routinely identified either in designated multi-wavelength survey fields, with sky coverage of a few square degrees, such as COSMOS \citep[e.g., ][]{Alexis2006} and XMM-LSS \citep[e.g., ][]{Bremer2006}, or larger area serendipitous archival surveys, e.g., the XCS \citep{Stanford06} and XDCP \citep{Mullis05}.


However, significant progress is still required to fully characterize the high redshift ($z\!\ga\!0.9$) cluster population, which has been limited to few known and well studied test objects so far \citep[e.g., ][]{Rosati2004,Hashimoto2005}. 
Detailed studies have recently revealed  large-scale structure (LSS) associated with several of the known bona fide X-ray clusters in the Lynx field at $z=1.27$ \citep{Nakata2005} and  for RDCS\,J1252.9-2927 at $z=1.24$ \citep{Tanaka2007}. 

The paper is organized as follows: Sect.\,\ref{s2_obs} describes the observations; Sect.\,\ref{s3_disc} discusses the cluster properties and its associated LSS; 
we conclude in Sect.\,\ref{s4_concl}.  
We assume a $\Lambda$CDM cosmology with $\Omega_\mathrm{m}\!=\!0.3$, $\Omega_{\Lambda}\!=\!0.7$, and $h\!=\!0.7$. 
All reported magnitudes are given in the Vega system.
At redshift $z\!=\!0.95$, the lookback time is 7.5\,Gyr and one arcsecond corresponds to a physical comoving scale of 7.9\,kpc. 




\section{Observations}
\label{s2_obs}

The X-ray data reduction and follow-up observations discussed here are part of the XMM-Newton Distant Cluster Project (XDCP), a serendipitous archival X-ray survey for very distant clusters of galaxies.
The survey strategy focusses on the spectroscopic confirmation of cluster candidates with estimated redshifts of $z\!>\!0.9$. The cluster XMMU\,J0104.4-0630 was an early $z\!\sim\!1$ candidate identified within the scope of a pilot study \citep[][]{HxB05}.

\subsection{X-ray Data}

We selected the galaxy cluster XMMU\,J0104.4-0630 as a serendipitous extended X-ray source in a high-galactic latitude archival field ($b\!=\!+69$\degr ; nominal exposure time: 26.7\,ksec; OBSID: 0112650401; here field 1). A second, partially overlapping XMM-Newton observation (nominal exposure time: 24.9\,ksec; OBSID: 0112650501; here field 2) extends the X-ray coverage to the south, providing moderately deep X-ray data over a field of about $25\arcmin \times 45\arcmin$.

We reduced the data with \texttt{SAS} v6.5;  the updated SAS versions 7.x are not expected to have a significant impact on our results. High background periods were excluded by applying a two-step flare cleaning procedure first in the hard (10-14\,keV) and subsequently in the 0.3-10\,keV band following \citet{Pratt2002}. 
The remaining clean exposure times for field 1 (field 2) are 15.1\,ksec (13.7) for the EPIC PN camera; 23.2\,ksec (19.7) for MOS1; and 23.5\,ksec (21.2) for MOS2.
We created images in different bands for the PN and MOS detectors from the cleaned event lists and later combined for the overlapping sky region of fields 1 and 2. 
A total of six extended X-ray sources were detected in the individual fields with the \texttt{SAS} tasks \texttt{eboxdetect} and \texttt{emldetect}. 
This letter focusses on two of the sources with available follow-up data, the main galaxy cluster XMMU\,J0104.4-0630 (here cluster\,A), and a second system  XMMU\,J0104.1-0635 at a projected distance of 6.4\arcmin \ to the SW (here cluster\,B). 
We detected cluster\,A at an off-axis angle of 5.5\arcmin \ in field\,1 with an aperture corrected, unabsorbed flux of $(1.7  \pm 0.3)\times 10^{-14}$\,erg\,s$^{-1}$\,cm$^{-2}$ \ in the 0.5--2.0\,keV energy band, using  $N_{\mathrm{H}}\!=\!5.08\times 10^{20}$\,cm$^{-2}$ \citep{Dickey1990a}.
The estimated core radius for a $\beta\!=\!2/3$ model is ($18.8 \pm 0.8$)\arcsec \ corresponding to 150\,kpc at $z=0.95$. The flux of cluster\,B at 7.2\arcmin \ off-axis angle in field\,2  is $(1.2  \pm 0.4)\times 10^{-14}$\,erg\,s$^{-1}$\,cm$^{-2}$, its estimated core radius of ($45 \pm 4$)\arcsec, corresponding to 340\,kpc, is significantly larger.
We checked the structure of both cluster sources and found that a potential flux contribution of single point sources outside the cluster cores is not more than 10\%.  



\subsection{Optical and Near-IR follow-up observations}

\begin{figure*}
   \sidecaption
   \centering
   \includegraphics[width=0.56\textwidth]{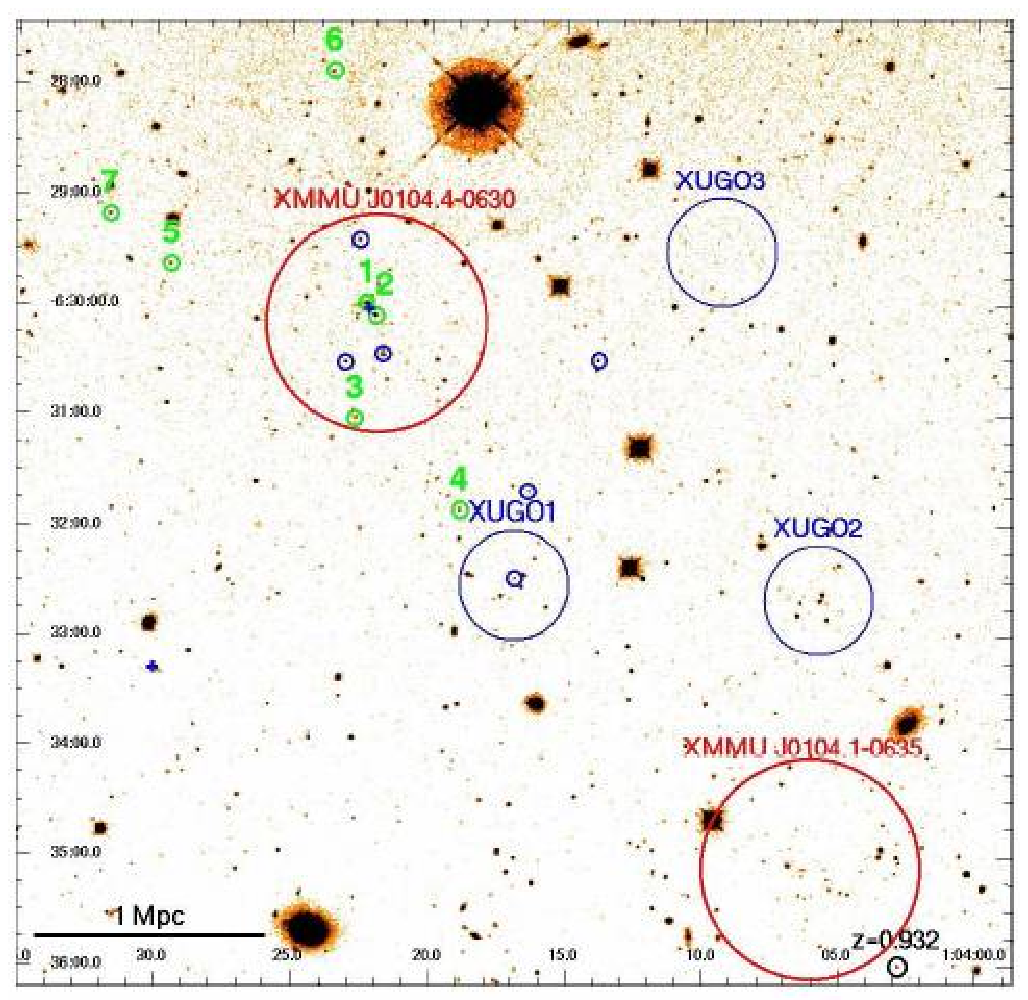}
   \includegraphics[width=0.425\textwidth]{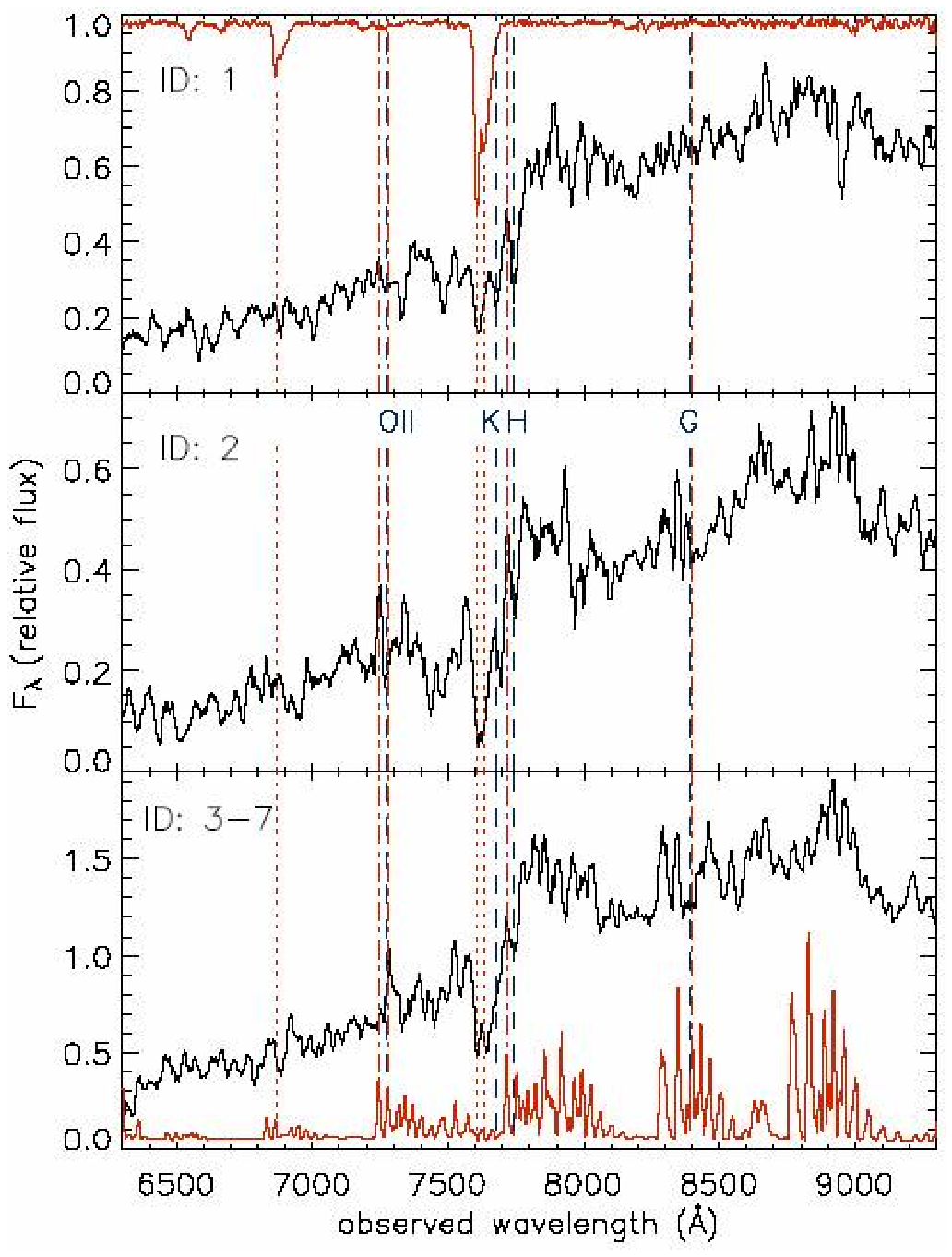}   
   \caption{{\em Left:} $9\arcmin \times 9\arcmin$ H-band image of the environment of XMMU\,J0104.4-0630. Spectroscopically confirmed cluster members are indicated by green circles with identification numbers, tentative cluster members are marked by small blue circles, radio sources by crosses, and the consistent literature redshift is shown in black. The cluster radio source is located in between the BCG (ID\,1) and the galaxy coincident with the X-ray center (ID\,2).
{\em Right:} VLT FORS\,2 spectra of the galaxies with secure and concordant redshifts 
giving the cluster redshift of $z\!=\!0.947\!\pm\!0.005$ as labelled in the left panel (smoothed with a 7\,pixel boxcar filter).
Flux units are arbitrary; sky emission lines and telluric absorption lines are shown in red at the bottom and top. 
The  two spectra from the top belong to the two brightest cluster galaxies in the center, the bottom panel displays the stacked spectrum of the galaxies with IDs 3--7.
}\label{f_spectra} \vfill
   \end{figure*}

\begin{figure*}[t]
   \sidecaption
   \centering
   \includegraphics[width=0.95\textwidth]{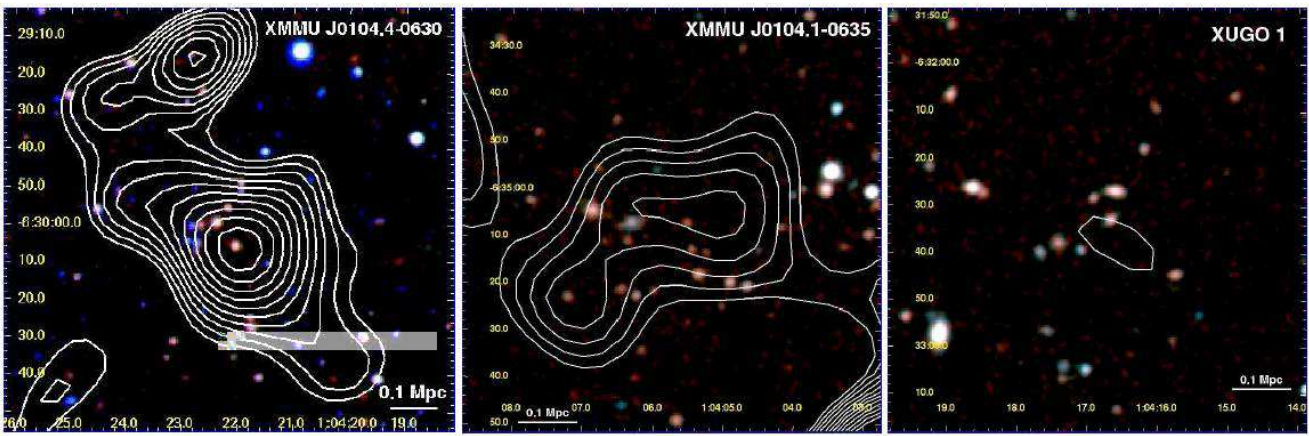}
   \caption{{\em Left:} $2\arcmin \times 2\arcmin$ R+z+H color image of XMMU\,J0104.4-0630 at $z\!=\!0.947$ with white X-ray contours overlaid. {\em Center:} $1.5\arcmin \times 1.5\arcmin$ z+H color composite of the second X-ray selected cluster XMMU\,J0104.1-0635 with a concordant redshift estimate. 
   {\em Right:} $1.5\arcmin \times 1.5\arcmin$ zoom on one of the X-ray undetected overdensities of bluer galaxies in the center of the field. 
              }
         \label{f_color_imas}
   \end{figure*}

\begin{figure*}
   \centering
   \includegraphics[width=\textwidth]{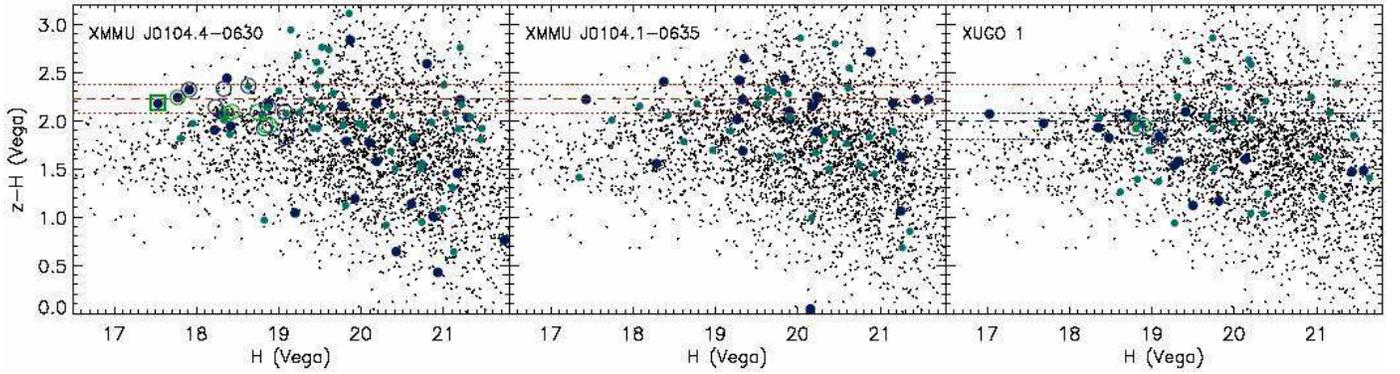}
   \caption{Corresponding z--H versus H color-magnitude diagrams of the three objects shown in Fig.\,\ref{f_color_imas}. Large circles indicate objects within a 30\arcsec radius from the center, smaller circles have cluster-centric distances of 0.5--1\arcmin, black dots represent all other objects in the field. {\em Left and center:}  
   CMDs of the two X-ray selected galaxy clusters with concordant red-sequence colors. The dashed red line indicates the predicted model color for passively evolving galaxies, the dotted red lines define a  $\pm 0.15$ magnitude interval around the model color representing the {\em red-galaxy population\/} in Fig.\,\ref{f_galoverdensities}. Spectroscopically confirmed cluster members of XMMU\,J0104.4-0630 are indicated in the left panel by open symbols. The BCG is marked by the green square, the six additional secure members by green circles, and the eight tentative cluster members by blue open circles.
{\em Right:} CMD of the galaxy overdensity between the two X-ray clusters with a bluer red-sequence color. 
The three galaxies there  with available spectroscopy are indicated by open circles.
The dashed blue line is shown at the color of the apparent object locus, the blue dotted lines define the color cuts for the {\em blue galaxy population\/} in Fig.\,\ref{f_galoverdensities}.   
                 }
         \label{f_CMD}
   \end{figure*}

The sky region around XMMU\,J0104.4-0630 was followed-up with R (1140s) and z-band (480s) imaging on 18 October 2003 with FORS\,2 (6.8\arcmin \ FoV) at the Very Large Telescope (VLT). We reobserved the target in H (1000s) and z (1800s) on 30 October 2006 in photometric conditions (1\arcsec \ seeing) with the NIR wide-field camera OMEGA2000 \citep{Bailer2000} at the Calar Alto 3.5m telescope with a larger 15.4\arcmin \ field-of-view, which also covered the second cluster XMMU\,J0104.1-0635 to the SW. 
The \texttt{Sextractor} \citep{Bertin1996} photometry in this larger field is calibrated to the Vega system using 2MASS point sources \citep{Cutri2003} in H, and designated Sloan Digital Sky Survey (SDSS) standard star observations \citep{Smith2002} in z. In both bands the limiting magnitude (50\% completeness) of $H_\mathrm{lim}\!\sim\!20.7$ and $z_\mathrm{lim}\!\sim\!22.8$ correspond to $m$*+1.6 for passively evolving galaxies at the cluster redshift.
Figure\,\ref{f_spectra} provides an overview of the field, Fig.\,\ref{f_color_imas} displays the R+z+H color composite image of XMMU\,J0104.4-0630, the z+H images of XMMU\,J0104.1-0635, and the region halfway between the clusters. The corresponding z--H color magnitude diagrams (CMDs) are shown in Fig.\,\ref{f_CMD}.   

 



We obtained spectroscopic observations  on 2 November 2005 with a VLT-FORS\,2 MXU slit-mask centered on XMMU\,J0104.4-0630 for a total exposure time of 60\,minutes. 
The incomplete execution of the program originally scheduled for 3\,h  and the rather poor seeing conditions of about 2\arcsec \  resulted in  a data quality that only allowed a redshift determination for about half the  targeted sources. 
However, seven galaxies were found at the same redshift and could thus be identified as 
secure cluster members (Fig.\,\ref{f_spectra}, right panel), yielding a cluster redshift for XMMU\,J0104.4-0630 of $z\!=\!0.947\!\pm\!0.005$.
Eight additional galaxies (blue circles in Fig.\,\ref{f_spectra}) are classified as tentative members, with indications of the D4000 break at the expected position but significant contamination from telluric absorption and sky emission lines.

\begin{table*}[t]    
\caption{Properties of the two newly-discovered X-ray clusters.}
\label{table:results}

\centering
\begin{tabular}{ c c c c c c c c c }
\hline \hline

ID & Name & RA & DEC	& $z$  & $f_{\mathrm{X}}$(0.5--2.0\,keV) & $r_{\mathrm{c}}$  &  $L_{\mathrm{X}}$(0.5--2.0\,keV)  \\ 

 & &  Eq. 2000 &  Eq. 2000	&    & erg\,s$^{-1}$cm$^{-2}$ & \arcsec &  erg\,s$^{-1}$  \\ 
 
\hline


A & XMMU\,J0104.4-0630 &  16.09148 & -6.50158 & $0.947\pm 0.005$   & $(1.7  \pm 0.3)\times 10^{-14}$ &  $18.8 \pm 0.8$  & $(6.4\pm 1.3) \times 10^{43}$  \\
B & XMMU\,J0104.1-0635 &  16.02967  & -6.58934 &  $0.95\pm 0.05$    & $(1.2  \pm 0.4)\times 10^{-14}$       &  $45 \pm 4$      &  $(4.4\pm 1.4) \times 10^{43}$  \\

\hline
\end{tabular}
\end{table*}

   \begin{figure*}
   \sidecaption
   \centering
   \includegraphics[width=0.664\textwidth]{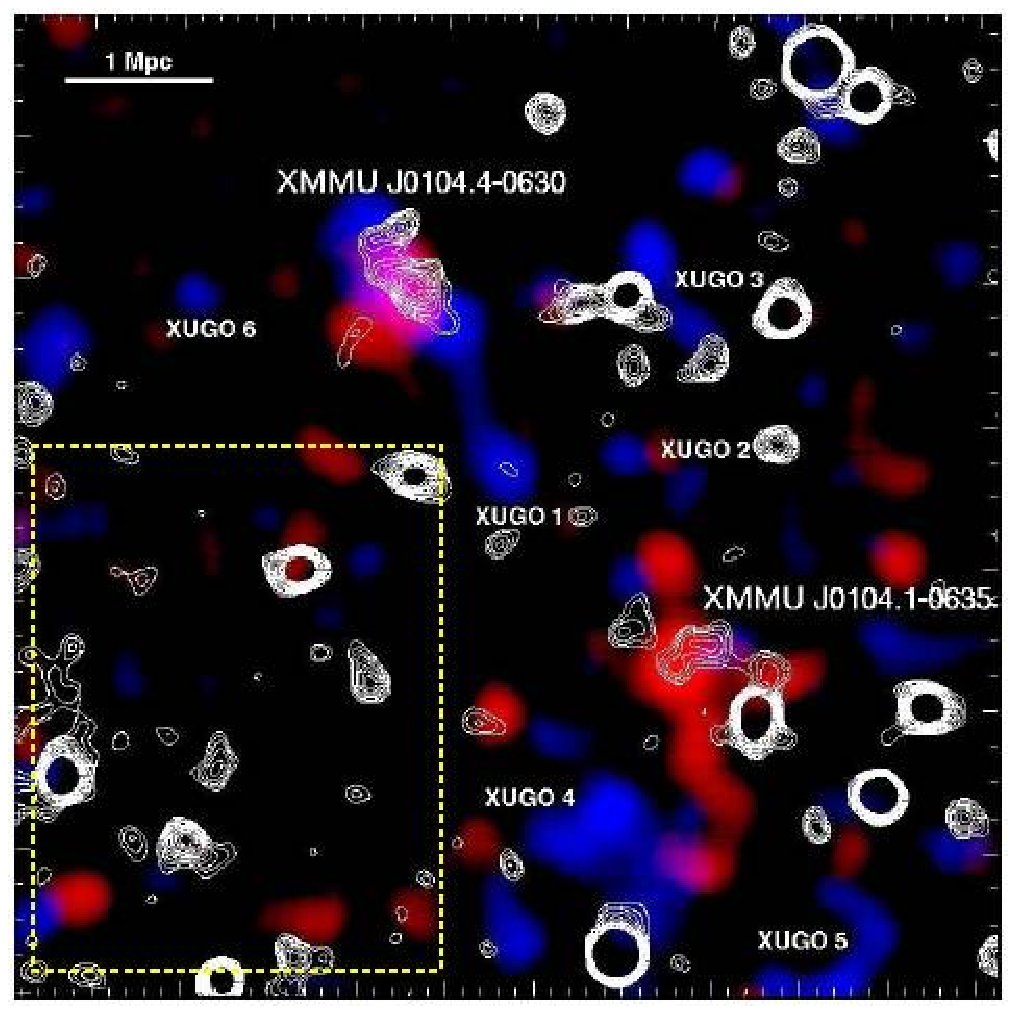}
   \includegraphics[width=0.331\textwidth]{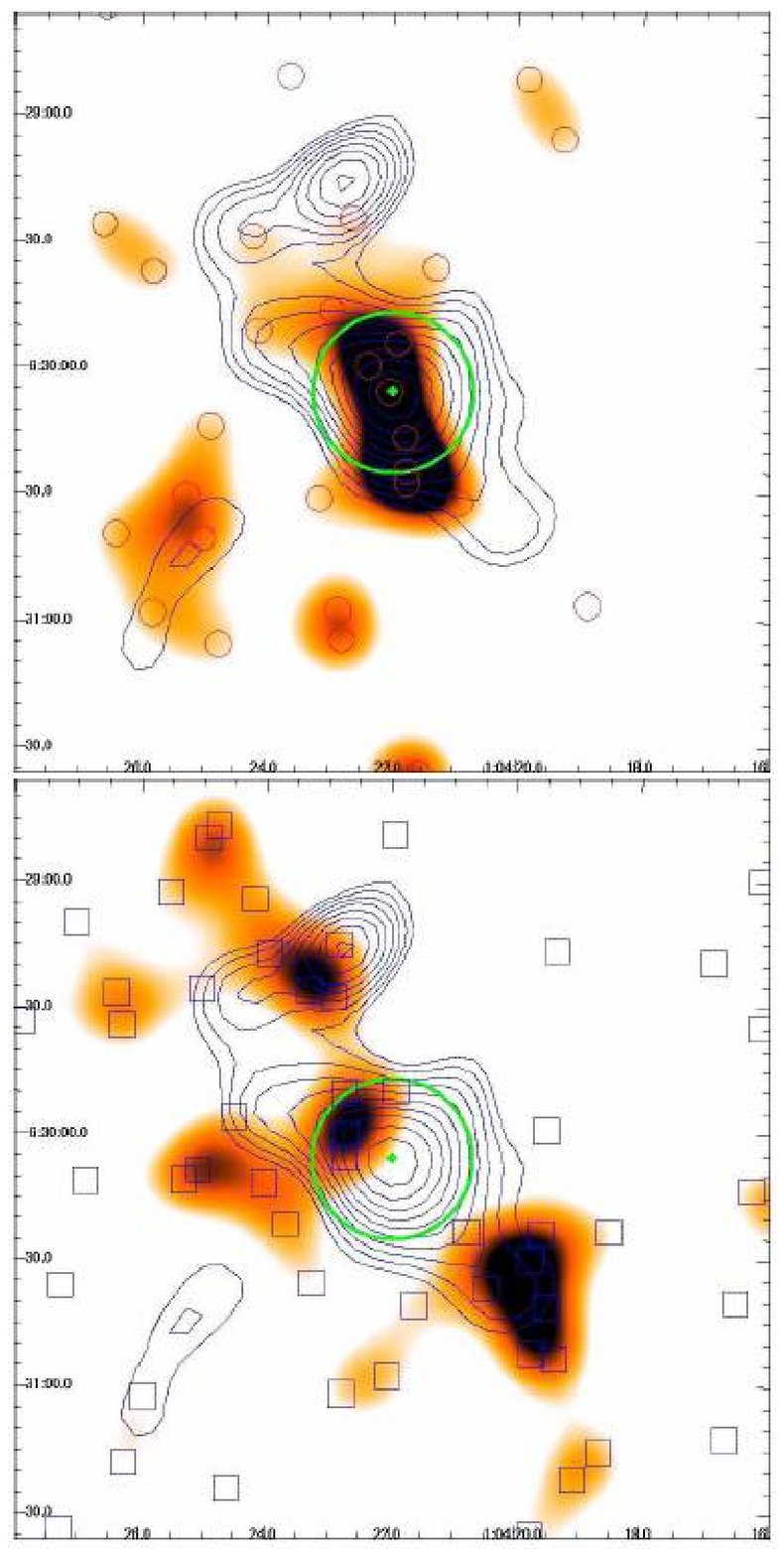}
   \caption{
{\em Left:} $14\arcmin\!\times\!14\arcmin$ field-of-view of galaxy overdensities smoothed with a 150\,kpc-kernel  of the {\em red\/} and {\em blue\/} galaxy populations (as defined in Fig.\,\ref{f_CMD}) encoded with the corresponding color; X-ray contours are overlaid in white. The displayed cuts for both populations span a logarithmic scale of $1.5-5$ sigma  relative to the lower left quadrant (yellow box), which served as the control field. Labels show the locations of the two X-ray selected galaxy clusters corresponding to 7-sigma peaks in the red population, and optically selected 3-6 sigma peaks (XUGO 1--6) in the blue population.   
{\em Right:} $3\arcmin\!\times\!3\arcmin$ zoom on the red (top) and blue (bottom) galaxy populations of XMMU\,J0104.4-0630 with blue X-ray contours overlaid. Overdensities are smoothed with a 75\,kpc-kernel and displayed with linear cuts of 2--5 sigma. The locations of individual red and blue objects are indicated by small symbols, the green cross marks the X-ray center, and the circle corresponds to the 150\,kpc core radius of the cluster.
}
         \label{f_galoverdensities}
   \end{figure*}



\section{Discussion}
\label{s3_disc}
\subsection{Cluster XMMU\,J0104.4-0630}




Cluster XMMU\,J0104.4-0630 (Fig.\,\ref{f_color_imas}, left),  with a 0.5--2.0\,keV luminosity of $L_\mathrm{x}\!=\!(6.4 \pm 1.3) \times  10^{43}$\,erg\,s$^{-1}$, is an intermediate-mass cluster of $M_{500}\!\sim\!1.1 \times 10^{14}\,M_{\sun}$ and $T_{\mathrm{X}}\!\sim\!3$\,keV, based on scaling relations following \citet{Alexis2006}. 
The system has a compact and fairly regular X-ray appearance with slight extensions to the NE and SW. 


The cluster core is dominated by early-type red galaxies as in low redshift clusters. The X-ray center coincides with the second brightest galaxy (ID 2 in Fig.\,\ref{f_spectra}). The brightest cluster galaxy (BCG) and the densest part of the cluster core exhibit an offset of about 10\arcsec \ to the NE. 
The color of the cluster red-sequence of z--H$\simeq$2.2 (Vega) is fully consistent with model predictions \citep{Fioc1997} of solar metallicity passively evolving galaxies with formation redshift $z_\mathrm{f}\!\sim\!5$ (Fig.\ref{f_CMD}). 


The cluster center has been identified  as the position of a 1.4\,GHz radio source with a flux of $11.9 \pm 1.0$\,mJy in the NRAO VLA Sky Survey (NVSS) \citep{Condon1998}, which is likely to be  attributed to a radio galaxy with a radio power $P_{1.4\,\mathrm{GHz}}\!\sim\!5\!\times\!10^{25}$\,W\,Hz$^{-1}$. 
The positional error circle of a few arcseconds radius includes the two brightest cluster galaxies. The fairly strong radio emission could be an indication of cooling core activity in the cluster center. The peak of the X-ray emission would then suggest that the cooling is actually associated with the second brightest cluster galaxy and not the BCG. In any case, radio emission of this order in high-z clusters is a prime concern for the cluster selection of upcoming SZ-surveys \citep{Lin2007} and requires detailed further studies.





\subsection{Associated large-scale structure environment}

The CMD of the second X-ray selected cluster  XMMU\,J0104.1-0635
reveals a color fully consistent 
with the same redshift as
the main cluster A, but with a fainter galaxy population. The overdensity of red galaxies (Fig.\,\ref{f_galoverdensities}) selected with a (red) color-cut of $2.23 \pm 0.15$ 
 shows a 7-sigma peak (6.5 sigma for cluster A) relative to the mean density (1.6 per square arcminute) and standard deviation (1.4)
in the 45 square arcminute control field in the lower left quadrant. We thus assign a photometric red-sequence redshift of $z\!=0.95\!\pm\!0.05$ to cluster B.
 We propose that clusters A and B are physically associated with each other, forming a double cluster with  projected separation in the plane of the sky of 3\,Mpc. 
   A literature redshift of $z\!=\!0.932$ from the XMM-Newton Serendipitous Survey \citep{Barcons2002} for a BL\,Lac object 75\arcsec \ to the SW of cluster B
  supports the idea of a large-scale structure filament along the axis of the two systems.  
  Currently, only two X-ray luminous double clusters at higher redshifts are reported in the literature \citep{Hashimoto2005,Nakata2005}.

The center of the proposed 3\,Mpc-scale cluster-cluster-bridge is marked by an additional significant overdensity of slightly bluer galaxies, with very weak X-ray emission well below the detection threshold for extended sources. 
The color composite and CMD of this X-ray 
undetected galaxy overdensity (XUGO) are shown in the right panels of Figs.\,\ref{f_color_imas} \& \ref{f_CMD} revealing a tight red-sequence at $\textrm{z}\!-\!\textrm{H}\!=\!2.0$, i.e. at
 a $\sim\!0.25$\,mag bluer color. 
Based on this color and assuming passively evolving early-type galaxies, the optically selected overdensity  would be consistent with a group or low-mass cluster at $z\!\sim\!0.7$. 
In order to investigate the nature of this system, we set a second (blue) adjoint color-cut at $1.8\!\le\!\textrm{z}-\textrm{H}\!<2.08$, corresponding to the blue dotted lines in Fig.\,\ref{f_CMD} with an average object density of 2.7 per square arcminute. The spatial distribution of  color selected objects was smoothed with a Gaussian kernel of 150\,kpc physical scale (19\arcsec), the approximate core radius of cluster A. The resulting overdensity plot of the $14\arcmin\!\times\!14\arcmin$ field-of-view is shown in Fig.\,\ref{f_galoverdensities}.
The two brightest peaks in the red galaxy population mark the X-ray selected clusters and are well-centered on the diffuse X-ray emission. On the other hand, the optically selected  system (XUGO\,1) corresponds to a 4.5-sigma peak of the bluer population. Figure\,\ref{f_galoverdensities} shows that this population can be traced to the outskirts of the main cluster A, where blue 5-sigma peaks to the NE and SW enclose  the cluster center along the axis laid out by the X-ray morphology. 
The density drop in the very center of cluster A in conjunction  with the rising  red overdensity (Fig.\,\ref{f_galoverdensities}, top right)
suggests that this is a direct consequence of the proposed scenario of delayed star formation quenching in lower density environments  
\citep[e.g., ][]{Thomas2005,Cuc2006}. Under the burst model assumption, the observed  0.15--0.3\,mag  z--H color shift toward the blue is consistent with an age difference of the stellar populations of 1.2--2.2\,Gyr. Systematic blue shifts of the S0 population have been observed in several clusters at lower redshift  
\citep[e.g., ][]{vanDokkum1998, Abraham1996} and at $z\!\sim\!1.1$ \citep{Mei2006}.  
The observed {\em blue\/} population in Fig.\,\ref{f_galoverdensities}, which includes the spectroscopic cluster members with IDs 3, 4, 6 and two additional tentative  members, could hence be interpreted as an evolving S0 population.  


On the other hand, the {\em blue\/} galaxies surrounding  XMMU\,J0104.4-0630 appear to be extending all to way to XUGO1.
Assuming that XUGO1 is at the same redshift as clusters A and B, as suggested by the geometric alignment, then its observed red-sequence would 
be significantly bluer (at the same bright-end magnitudes). 
Since such a red-sequence  blueing in LSS filaments has never been observed, XUGO1 and the other five X-ray undetected blue galaxy overdensities on the 3--5\,$\sigma$ level (XUGO2--6) could well be  foreground groups. Firm conclusions on physically associated filaments of XMMU\,J0104.4-0630
will only be possible with additional spectroscopic observations of the field.

\section{Summary and Conclusions}
\label{s4_concl}
   \begin{enumerate}
      \item We have presented details of the newly-discovered X-ray luminous cluster of galaxies XMMU\,J0104.4-0630 at redshift $z\!=\!0.947\!\pm\!0.005$.
       The compact, intermediate mass cluster is found to be in an evolved state and hosts a strong central radio source,  which could be a sign of ongoing cooling core activity.  
       \item The cluster shows a pronounced stratification of galaxy populations. The spatial distribution of the red-sequence population (4 spectroscopic members) of early-type galaxies coincides well with the X-ray emission, whereas a significantly bluer population (3 spectroscopic members) dominates beyond 1--2 core radii from the center, suggesting a cluster environment-driven effect of differential galaxy evolution, e.g.,  delayed star formation quenching in the outskirts.
       
      \item The second  X-ray selected cluster, XMMU\,J0104.1-0635,  6.4\arcmin \ to the SW with  a consistent red-sequence color likely forms a cluster-cluster bridge with the main system. XMMU\,J0104.1-0635 seems to be in a less evolved state exhibiting a less compact and more irregular X-ray morphology and spatial galaxy distribution. 
      \item Further spectroscopic LSS studies in this field have the potential to reconstruct the cosmic web structure at $z\,\sim\,1$ on the 10\,Mpc scale and investigate environmental effects of galaxy evolution at a lookback time of 7.5\,Gyr.


   \end{enumerate}

\begin{acknowledgements}
The XMM-Newton project is an ESA Science Mission with instruments
and contributions directly funded by ESA Member States and the
USA (NASA). The XMM-Newton project is supported by the
Bundesministerium f\"ur Wirtschaft und Technologie/Deutsches Zentrum
f\"ur Luft- und Raumfahrt (BMWI/DLR, FKZ 50 OX 0001), the Max-Planck
Society and the Heidenhain-Stiftung.
This research has made use of the NASA/IPAC Extragalactic Database (NED) which is operated by the Jet Propulsion Laboratory, California Institute of Technology, under contract with the National Aeronautics and Space Administration.
GL is supported by DLR under contract  500OX201, JK and JSS are supported by DFG under contracts Schw536/24-1 and BO702/16-2, respectively. We acknowledge
additional support from the excellence cluster {\em Universe EXC153}.
RF would like to thank Gabriel Pratt, Daniele Pierini, and Bianca Poggianti for helpful discussions and Filiberto Braglia, Aurora Simeonescu, and Martin M\"uhlegger for supporting the observations.
\end{acknowledgements}

\end{document}